\documentclass[aps,showpacs,twocolumn]{revtex4}
\usepackage{epsfig}
\usepackage{dcolumn}
\usepackage{bm}

\begin{document}
\author{Youchang Yang$^{a,b}$ and Jialun Ping$^a$}
\email{jlping@njnu.edu.cn}

\affiliation{$^a$Department of Physics, Nanjing Normal University,
Nanjing 210097, P. R. China\\$^b$Department of Physics, Zunyi Normal
College, Zunyi 563002, P. R. China}

\title{Dynamical study of the $X$(3915) as a molecular $D^*\bar{D^*}$ state in a quark model }

\begin{abstract}
Considering the coupling of color $1 \otimes 1$ and  $8 \otimes 8$
structures, we calculate the energy of the newly observed $X$(3915) as
$S-$wave $D^*\bar{D^*}$ state in the Bhaduri, Cohler, and Nogami
quark model by the Gaussian Expansion Method. Due to the color
coupling, the bound state of $D^*\bar{D^*}$ with $J^{PC}=0^{++}$ is
found, which is well consonant with the experimental data of the
$X$(3915). The bound state of $B^*\bar{B^*}$ with $J^{PC}=0^{++}$
and $2^{++}$ are also predicted in this work.
\end{abstract}

\maketitle

\section{Introduction}
Very recently,
Belle Collaboration reported the newest charmonium like state, the $X(3915)$, which is
observed in $\gamma\gamma \to \omega J/\psi$ with a statistical significance of
7.7$\sigma$~\cite{belle:x3915,Olsen0909,CZYuan0910,Zupanc0910,Gordfrey0910}.
It has the mass $M=3915\pm3\pm2$~MeV and width $\Gamma=17\pm 10\pm3$~MeV. Belle
Collaboration determined the $X(3915)$ production rate
\begin{eqnarray}
\Gamma_{\gamma \gamma}(X(3915)) {\cal B}(X(3915) \to \omega
J/\psi)=61 \pm 17 \pm 8~ {\rm eV} \label{Xgamma1}
\end{eqnarray}
or
\begin{eqnarray}
\Gamma_{\gamma \gamma}(X(3915)) {\cal B}(X(3915) \to \omega
J/\psi)=18 \pm 5 \pm 2~ {\rm eV},\label{Xgamma2}
\end{eqnarray}
for $J^P=0^+$ or $2^+$, respectively. The $X(3915)$ is similar to
the previously discovered $Y(3940)$ by Belle and confirmed by
BaBar Collaboration in the process $B \to KY(3940),~ Y(3940)\to\omega
J/\psi$~\cite{Abe:2004zs,Aubert:2007vj}. The mass and width of the $Y(3940)$ obtained by
Belle are: $M = 3943 \pm 11 ({\rm stat}) \pm 13
({\rm syst})$ MeV, $\Gamma = 87 \pm 22 ({\rm stat}) \pm 26 ({\rm
syst})$ MeV~\cite{Abe:2004zs}. However, BaBar obtained
$M = 3914.6^{+3.8}_{-3.4} ({\rm stat}) \pm 2.0 ({\rm
syst})$ MeV, $\Gamma = 34^{+12}_{-8} ({\rm stat}) \pm 5 ({\rm
syst})$ MeV~\cite{Aubert:2007vj}. The difference of the mass and width of $Y$(3940)
might be the attribute to the large data sample used by BaBar (350
fb$^{-1}$ compared to Belle's 253 fb$^{-1}$), which enabled them to
use smaller $\omega J/\psi$ mass bins in their analysis
\cite{Olsen0909,CZYuan0910}. The $Y(3940)$ and $X(3915)$ have same
mass and width, and are both seen in $ \omega J/\psi$. Hence they
might be same states pointed out in
Refs.\cite{belle:x3915,Olsen0909,CZYuan0910,Zupanc0910,Gordfrey0910}.

The $X(3915)$ as a conventional $c\bar c$ charmonium interpretation
is disfavored \cite{belle:x3915,Olsen0909,CZYuan0910}. The mass of
this state is well above the threshold of $D\bar{D}$ or $D\bar{D}^*$
\cite{PDG}, hence open charm decay modes would dominate, while the
$\omega J/\psi$ decay rates are essentially negligible, which is
same as $Y(3940)$~\cite{Eichten:2007qx}. However, according to the
production rate obtained by Belle, Eqs.(\ref{Xgamma1}) and (\ref{Xgamma2}),
we obtain the $\Gamma_{\omega
J/\psi}\sim\mathcal{O}$ (1 MeV) if we assume $\Gamma_{\gamma
\gamma}\sim\mathcal{O}$ (1$\sim$ 2 keV) which is typical for an
excited charmonium state
\cite{Ebert:gamma,Munz:gamma,Badalian,chaoprd80}. $\Gamma_{\omega
J/\psi}\sim\mathcal{O}$ (1 MeV) is an order of magnitude higher than
typical rates between known charmonium states, which imply that
there is intricate structure for the $X(3915)$. In our previous
study \cite{yang:X3915}, the quantum number and mass of
$\chi_0(2^3P_0)$ is compatible with the $X(3915)$. However, the
calculated strong decay width is much larger than experimental data.
Hence, the assignment of $X(3915)$ to the charmonium
$\chi_0(2^3P_0)$ state is very unlikely.

In the previous work, Liu~\cite{Liuxiang:Y3940epjc,Liuxiang:Y3940}
suggested $Y(3940)$ is
probably a molecular $D^*\bar{D^*}$ states with $J^{PC}=0^{++}$ or
$J^{PC}=2^{++}$ in a meson-exchange model. Assuming the $D^*\bar{D^*}$
bound-state structure, Branz \cite{Branz:Y3940} studied the strong
$Y(3940)\to \omega J/\psi$ and radiative $Y(3940) \to \gamma\gamma$
decay widths in a phenomenological Lagrangian approach. Their
results are roughly compatible with the experimental data about
$Y(3940)$. By the QCD sum rules, Zhang \cite{JianRongZhang:Y3940}
also obtained the mass $M=3.91\pm0.11$ GeV for $D^*\bar{D^*}$, which
is consistent with $Y(3940)$ reported by BaBar. Using SU(3) chiral
quark model and solving the resonating group method equation,
Liu and Zhang~\cite{LiuYanRui:Y3940} also studied molecular $D^*\bar{D^*}$
states with color singlet meson-meson structure, the bound
state of $D^*\bar{D^*}$ may appear if the vector meson exchange
between light quark and antiquark is considered.

Although at present the data for the $X$(3915) and $Y(3940)$ are
still preliminary, and whether the $X(3915)$ is the same as
the $Y(3940)$ needs to be identified, it is worthwhile to discuss its
possible assignments, especially in view of the great potential of
finding new particles. We assume the $X$(3915) and $Y(3940)$ as same
state in the following study.
In this work we study the $X(1915)$ with molecular $D^*\bar{D^*}$
structure by taking into account of the coupling of different color structures
in the Bhaduri, Cohler and Nogami (BCN) quark model, since the color
singlet tetraquark state $(c\bar{q})^*(\bar{c}q)^*$ can be consisted
of colorless or
colored $(c\bar{q})^*$ and $(\bar{c}q)^*$ two-quark mesons. The role of the
\textit{\textbf{hidden color}} in $qq-\bar{q}\bar{q}$ ($q=u,d$) has been
discussed by Brink and Stancu \cite{Brink:hiddencolor}, and then extended
to study other four-quark systems~\cite{Brink,Hogaasen,PRD114023}.

\begin{figure}[htb]
\includegraphics{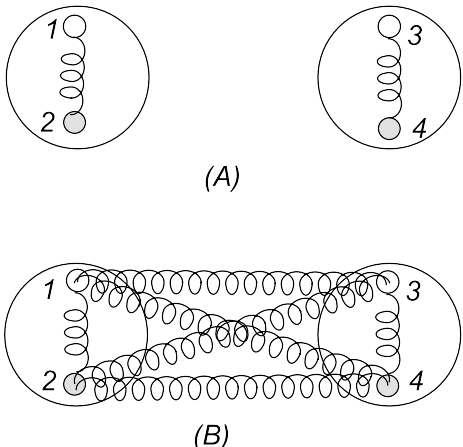}
\vspace*{-0.5cm}
 \caption{The molecule bound state by a one-gluon exchange interaction
in color structures $1 \otimes 1$ (A) and $8 \otimes 8$ (B) in the BCN
quark model.
The darkened and open circles represent quarks and antiquarks,
respectively.}\label{BCNmolecule}
\end{figure}

The BCN quark model was proposed by Bhaduri, Cohler and Nogami~\cite{Bhduri}.
In this model, the interaction between quark pair is
composed of linear color confinement, color Coulomb and color
magnetic interaction from one-gluon-exchange. This model has been
widely used to study $q\bar{q}$ meson, $qqq$ baryon \cite{Bhduri1}
and four-quark system \cite{Bhduri2, Brink, Janc1, Janc2} range from
light quark $u,~d$ to heavy quark $b$ with same set of parameters.
The molecular
state bound by a one-gluon exchange interaction in the BCN quark
model can be depicted in Fig. \ref{BCNmolecule}. One-gluon exchange
interaction exist between pair of quark-quark or quark-antiquark in
color structure $8 \otimes 8$, while no interaction between meson-meson in
color structure $1 \otimes 1$. Hence there is no bound state in Fig.\ref{BCNmolecule}(A).
However, the bound state is possible within hidden color
channel (color structure $8 \otimes 8$, Fig.\ref{BCNmolecule}(B)).

The numerical method, which is able to provide reliable solutions, is
very important for calculating the energy of few-body systems. In
this work, the spectra of normal and exotic mesons are obtained by
solving the Sch\"{o}dinger equation using a variational method:
Gaussian Expression Method (GEM), which is a high precision
numerical method for few body system. The detail about GEM can be
seen in Refs. \cite{GEM,PRD114023}.

The paper is organized as follows. In the next section we show the
Hamiltonian and parameters of BCN quark model. Section \ref{WF} is
devoted to discuss the wave function of $X(3915)$ and similar state
$X_B$ with $c$ quark replaced by $b$ quark. In Section \ref{result}, we present and
analyze the results obtained by our calculation. Finally, the
summary of the present work is given in the last section.

\section{Hamiltonian}
The Hamiltonian of BCN model takes the form,
\begin{equation}
H=\sum_{i=1}^4\left(m_i+\frac{\mathbf{p}_i^2}{2m_i}\right)-T_{c.m.}+\sum_{j>i=1}^4
(V_{ij}^C+V_{ij}^G)\label{bcnh}
\end{equation}
with
\begin{eqnarray}
&&V_{ij}^G=\frac{\mathbf{\lambda}_i^c\cdot\mathbf{\lambda}^c_j}{4}
\left(\frac{\kappa}{r_{ij}}-\frac{\kappa}{m_im_j}\frac{e^{-r_{ij}/r_0}}{r_0^2r_{ij}}\mathbf{\sigma}_i\cdot\mathbf{\sigma}_j\right),\\
&&V_{ij}^{C}=\mathbf{\lambda}^c_{i}\cdot \mathbf{
\lambda}^c_{j}~(-a_{c}r_{ij}- \Delta).\label{bcnv}
\end{eqnarray}
where $r_{ij}=|\mathbf{r}_i-\mathbf{r}_j|$ and $T_{c.m.}$ is the
kinetic energy of central-mass of whole system. $\mathbf{\sigma}$,
$\mathbf{\lambda}$ are the SU(2) Pauli and the SU(3)
Gell-Mann matrices, respectively. The $\mathbf{\lambda}~$ should be
replaced by $-\mathbf{\lambda}^{*}$ for the antiquark. The universal
parameters of this model \cite{Bhduri,Janc1}  are listed in Table
\ref{parameters}.

\begin{ruledtabular}
\begin{table}[htb]
\caption{Parameters of the BCN model. }\label{parameters}
\begin{tabular}{clccc}
              & $m_{u,d}$ (MeV)       &337  \\
 Quark masses & $m_s$ (MeV)           &600  \\
              & $m_c$ (MeV)            &1870  \\
              & $m_b$ (MeV)            &5259  \\ \hline
  Confinement & $a_c$ (MeV fm$^{-1}$) &176.738  \\
              & $\Delta$ (MeV)        &-171.25 \\ \hline
  OGE         & $\kappa $              & 0.390209\\
              & $r_{0}$ (fm)              &0.4545
\end{tabular}
\end{table}
\end{ruledtabular}

\section{wave function \label{WF}}
The total wave function of four-quark system can be written as,
\begin{equation}
\Psi^{I,I_z}_{J,J_z}=\left|\xi\right\rangle
\left|\eta\right\rangle^{II_z}\Phi_{JJ_z},\label{twave}
\end{equation}
with\[\Phi_{JJ_z}=\left[\left|\chi\right\rangle_{S}\otimes\left|\Phi\right\rangle_{L_T}\right]_{JJ_z}
\]
where $\left|\xi\right\rangle$, $\left|\eta\right\rangle^{I}$,
$\left|\chi\right\rangle^{S}$, $\left|\Phi\right\rangle_{L_{T}}$
represent color singlet, isospin with $I$, spin with $S$ and spacial
with angular momentum $L_T$ wave functions, respectively.

Due to the $X(3915)$ is observed in the invariant mass spectrum of
$J/\psi\omega$ in $\gamma \gamma \to J/\psi\omega$
\cite{belle:x3915}, the flavor wave functions of the possible
molecular states is
\begin{equation}
X(3915)=\frac{1}{\sqrt{2}}\left(\bar{D}^{0*}D^{0*}+D^{-*}D^{+*}\right)
.
\end{equation}
The similar state composed of $b$ quark is
\begin{equation}
X_B=\frac{1}{\sqrt{2}}\left(\bar{B}^{0*}B^{0*}+B^{-*}B^{+*}\right) .
\end{equation}

\begin{figure}[htb]
\includegraphics{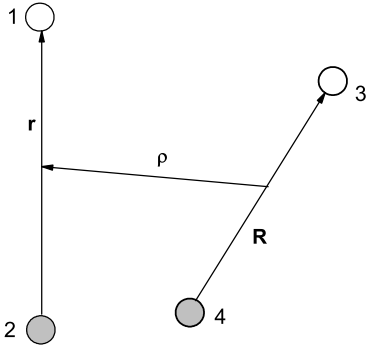}
\vspace*{-0.5cm}
 \caption{The relative
coordinate of a molecular state. The darkened and open circles
represent quarks and antiquarks, respectively.}\label{jacobi}
\end{figure}

The spatial structure of the above states are pictured in Fig.
\ref{jacobi}. The relative coordinates are defined as following,
\begin{eqnarray}
&&\mathbf{r}=\mathbf{r}_1-\mathbf{r}_2,\ ~~~~ \mathbf{R}=\mathbf{r}_3-\mathbf{r}_4,\\
&&\mbox{\boldmath$\rho$}=\frac{m_1\mathbf{r}_1+m_2\mathbf{r}_2}{m_1+m_2}
-\frac{m_3\mathbf{r}_3+m_4\mathbf{r}_4}{m_3+m_4},
\end{eqnarray}
 and the coordinate of the center of mass is
\begin{equation}
\mathbf{R}_{cm}=\sum_{i=1}^4 m_i \mathbf{r}_i/\sum_{i=1}^4 m_i,
\end{equation}
where $m_i$ is the mass of the $i$th quark.

By GEM, the outer products of space and spin is
\begin{eqnarray}
\Phi_{JJ_z}&&=[[
\phi^G_{lm}(\mathbf{r})\chi_{s_1m_{s_1}}]_{J_1M_1}\nonumber\\
&&[\psi^G_{LM}(\mathbf{R})\chi_{s_2m_{s_2}}]_{J_2M_2}
]_{J_{12}M_{12}}\varphi^G_{\beta\gamma}(\mathbf{\rho})]_{JJ_z}.
\end{eqnarray}
The total angular momentum is 0 or 2 suggested by Belle
Collaboration \cite{belle:x3915}.

Three relative motion wave functions are written as,
\begin{eqnarray}
\phi^G_{lm}(\mathbf{r})=\sum_{n=1}^{n_{max}}c_nN_{nl}r^le^{-\nu_nr^2}Y_{lm}(\hat{\mathbf{r}})\label{gem1}
\end{eqnarray}
\begin{eqnarray}
\psi^G_{LM}(\mathbf{R})=\sum_{N=1}^{N_{max}}c_NN_{NL}R^Le^{-\zeta_NR^2}Y_{LM}(\hat{\mathbf{R}})\label{gem2}
\end{eqnarray}
\begin{eqnarray}
\varphi^G_{\beta\gamma}(\mbox{\boldmath$\rho$})=
\sum_{\alpha=1}^{\alpha_{max}}c_{\alpha}N_{\alpha\beta}\rho^{\beta}e^{-\omega_\alpha
\rho^2}Y_{\beta\gamma}(\hat{\mbox{\boldmath$\rho$}})\label{gem3}
\end{eqnarray}
Gaussian size parameters are taken as geometric progression
\begin{eqnarray}
\nu_{n}=\frac{1}{r^2_n},& r_n=r_1a^{n-1},&
a=\left(\frac{r_{n_{max}}}{r_1}\right)^{\frac{1}{n_{max}-1}}
\label{geo progress}
\end{eqnarray}
The expression of $\zeta_N,\omega_\alpha$ in Eqs.
(\ref{gem2}) - (\ref{gem3}) are similar to Eq. (\ref{geo progress}).

The physical state must be color singlet, so the color wave functions of the
possible molecular state can be constructed by the following color
structures,
 \begin{eqnarray}
\left|\xi_1\right\rangle=\left|\mathbf{1}_{12}\otimes\mathbf{1}_{34}\right\rangle,~~~
\left|\xi_2\right\rangle=\left|\mathbf{8}_{12}\otimes\mathbf{8}_{34}\right\rangle.
\label{color}
 \end{eqnarray}
Clearly, it includes a color singlet-singlet and a color octet-octet
state. The latter is called \textit{\textbf{hidden color}} states by
analogy to states which appear in the nucleon-nucleon problem
\cite{wang0903.4349}.

\section{Numerical results and discussion\label{result}}
Using GEM and the universal model parameters listed in Table
\ref{parameters}, we also calculate the mass of normal meson
composed of quark-antiquark by solving the Schr\"{o}dinger equation
\begin{eqnarray}
\left(H-E\right)\Psi^{I,I_z}_{J,J_z}=0 \label{schrodinger}
\end{eqnarray}
with Rayleigh-Ritz variational principle. The results of meson
spectra  are converged with the number of
gaussians $n_{max}=7$, and the relative distance of quark-antiquark
ranges from $0.1$ to $2$~fm which is determined by the convergence
properties of the energies and discussed in detail in Ref.
\cite{PRD114023}. The calculated results are listed in Table \ref{meson
spectrum} which are in good agreement with the experimental data.

\begin{center}
\begin{ruledtabular}
\begin{table}[htb]
\caption{The normal meson spectra (MeV) in BCN model. The last
column takes from PDG compilation~\cite{PDG}, and the
the ground state of bottom $\eta_b(1s)$ is observed by BABAR
Collaboration from the radiative transition $\Upsilon(3S)\rightarrow
\gamma\eta_b$ \cite{Grenier} }\label{meson spectrum}
\begin{tabular}{ccc}
Meson &BCN &Exp. \\ \hline
 $\pi$         &137.5    &139.57$\pm$0.00035\\
 K             &521.4    &493.677$\pm$0.016\\
 $\rho(770)$   &779.6    &775.49$\pm$0.34\\
 $K^*(892)$    &907      &896.00$\pm$0.25\\
 $\omega(782)$ &779.6    &782.65$\pm$0.12\\
 $\phi(1020)$  &1018.5   &1019.422$\pm$0.02\\
 $\eta_c(1s)$  &3040     &2980.3$\pm$1.2\\
 $J/\psi(1s)$  &3098     &3096.916$\pm$0.011\\
 $D^0$         &1886.7   &1864.84$\pm$0.17\\
 $D^*$         &2021.3   &2006.97$\pm$0.19\\
 $D_s$         &1997     &1968.49$\pm$0.34\\
 $D^*_s$       &2102.3   &2112.3$\pm$0.5\\
 $B^\pm$       &5302     &5279.15$\pm$0.31\\
 $B^0$         &5302     &5279.53$\pm$0.33\\
 $B^*$         &5351.5   &5325.1$\pm$0.5\\
 $B^0_s$       &5373.1   &5366.3$\pm$0.6\\
 $B^*_s$       &5414.5   &5412.8$\pm$1.3\\
 $\eta_b(1s)$  &9422.2   &9388.9$^{+3.1}_{-2.3}$(stat)\\
 $\Upsilon(1s)$ &9439.5  &9460.30$\pm$0.26
\end{tabular}
\end{table}
\end{ruledtabular}
\end{center}

Due to the threshold of four-quark system is governed by two
corresponding meson masses, so a good fit of meson spectra, with the
same parameters used in four-quark calculations, must be the most
important criterium \cite{Bhduri2,Brink,Janc1,weinstein,Manohar2}.
Therefore, the universal model parameters are also employed in this
work to study the possible molecular state of the $X$(3915) and
$X_B$.

Using GEM with the numbers of gaussians $\alpha=12, ~n=7,~N=7$, and
the magnitude of $\mathbf{z}$ ranges from 0.1 to 6~fm, and 0.1 to
2~fm for $\mathbf{x}$ and $\mathbf{y}$, respectively, in
Eqs.(\ref{gem1})-(\ref{gem3}), we obtain converged results for
the four-quark systems which also discussed in detail in Ref.
\cite{PRD114023}. By taking the coupling of color singlet and octet
into account, the total energy of the possible molecular state
mentioned above are listed in Table \ref{X3915}.
\begin{center}
\begin{ruledtabular}
\begin{table}[htb]
\caption{The energy of the possible molecular states.
E$_{th}$ and E$_{4q}$ are the threshold and
total energy of molecular state, respectively. The 'cc' stands
for channel coupling of color structures $1 \otimes 1$ and $8 \otimes 8$.}\label{X3915}
\begin{tabular}{ccc|ccc}
Molecular state &$J^{PC}$  & E$_{th}$ &\multicolumn{3}{c}{E$_{4q}$ (MeV)}
\\ & & (MeV) & $1 \otimes 1$ & $8 \otimes 8$ & cc \\ \hline
$D^*\bar{D}^*$ &$0^{++}$ &4042.6 &4043.6 &3941.9 &3912.1 \\
               &$2^{++}$ &4042.6 &4043.6 &4200.3 &4043.6 \\ \hline
$B^*\bar{B}^*$ &$0^{++}$ &10703  &10703.4 &10388.74 &10298.5 \\
               &$2^{++}$ &10703  &10703.4 &10609.83 &10535.1 \\
\end{tabular}
\end{table}
\end{ruledtabular}
\end{center}

The $D^*\bar{D}^*$ with $J^{PC}=0^{++}$ is bound and in good
agreement with the mass of the $X(3915)$ and Ref.
\cite{JianRongZhang:Y3940}, while the $J^{PC}=2^{++}$ with same
flavor wave function is unbound. The results can be
understood as follows. For convenience, The matrix elements of
color operator are given in Table \ref{colormatrix}, where the
$(i,j)$ $(j>i=1,2,3,4)$ are the indices of quark pairs.
The matrix elements of spin operator for interacting quark pairs
$(i,j)$ are
$((1,2),~(3,4),~(1,3),~(2,4),~(1,4),~(2,3))=(1,~1,~-2,~-2,~-2,~-2)$
and $(1,~1,~1,~1,~1,~1)$ for $J^{PC}=0^{++}$ and $2^{++}$,
respectively. For the color singlet-singlet channel, because of no 
interaction between $D^*$ and $\bar{D}^*$, the calculated energies 
of $D^*\bar{D}^*$ must converged to the the threshold, the sum of the
mass of $D^*$ and $\bar{D}^*$; the distances between two quarks (or antiquarks)
residented in the different color-singlets are much larger than that 
of quark and antiquark in the same color-singlet. The results are 
demonstrated in Table \ref{X3915} and \ref{radius}. 
For the hidden color channel, according to the
confinement interaction in Eq.(\ref{bcnv}) and color matrix element
listed in Table \ref{colormatrix}, the interactions between 
quark-antiquark pairs (1,2) and (3,4) are repulsive, anticonfinement
appears~\cite{anticonfine}. However, the other four pairwise interactions
are still attractive which govern the confinement of whole four-quark
system. The distance between quark-antiquark pair (1,2) or (3,4) is the competition 
result of the interaction between particle 1,2 (3,4) and that between 1,3, 1,4,
2,3 and 2,4. The distances are converged to values given in Table \ref{radius}
and the energy of the four-quark system is also stable (see Table~\ref{X3915}).
The color magnetic interaction are also important in the hidden color channel. 
For the state with $J^{PC}=0^{++}$, the contributions from color magnetic interaction
$-(\lambda^c_i \cdot\lambda^c_j)(\mathbf{\sigma}_i \cdot\mathbf{\sigma}_j)$ 
of each particle-pair are all attractive. Hence, in hidden color channel the energy of 
$D^*\bar{D}^*$ with $J^{PC}=0^{++}$ is lower than the state with $J^{PC}=2^{++}$,
where the contributions from color magnetic interaction for particle-pairs
(1,2) and (3,4) are attractive while that for other four particle-pair are
repulsive. The channel coupling increases the difference further, because the
cross matrix elements between color singlet-singlet and color octet-octet 
with $J^{PC}=0^{++}$ is two times larger than it with $J^{PC}=2^{++}$.
 
From the Table \ref{X3915}, one can see that the $B^*\bar{B}^*$ with $J^{PC}=0^{++}$
and $2^{++}$ are both bound states. The appearance of the state 
$B^*\bar{B}^*$ with $J^{PC}=2^{++}$ is due to the larger mass
of $b$ quark than the $c$ quark, the kinetic energy of former is lower
than the latter, the repulsive color magnetic interactions in
$B^*\bar{B}^*$ is also small (the interaction is inversely proportional
 to the square of quark mass) in hidden color channel which leads to
the energy below the threshold of $B^*\bar{B}^*$.

\begin{center}
\begin{ruledtabular}
\begin{table}[htb]
\caption{Color matrix elements $\langle
\xi_k|\mathbf{\lambda}_i^c\cdot\mathbf{\lambda}^c_j|\xi_l\rangle$
with $k,l=1,2$ and $j>i=1,2,3,4$. }\label{colormatrix}
\begin{center}
\begin{tabular}{c|cccccc}
$(i,j)=$ &(1,2)&(3,4) &(1,3)&(2,4)&(1,4)&(2,3) \\ \hline
 $(k,l)=(1,1)$ & $-\frac{16}{3}$&$-\frac{16}{3}$ &0&0&0&0 \\
 $(k,l)=(2,2)$ & $\frac{2}{3}$&$\frac{2}{3}$
   &$-\frac{4}{3}$&$-\frac{4}{3}$&$-\frac{14}{3}$&$-\frac{14}{3}$\\
 $(k,l)=(1,2)$ & 0&0 &$\sqrt{\frac{32}{9}}$&$\sqrt{\frac{32}{9}}$
 &$-\sqrt{\frac{32}{9}}$&$-\sqrt{\frac{32}{9}}$
 \\
\end{tabular}
\end{center}
\end{table}
\end{ruledtabular}
\end{center}

\begin{center}
\begin{ruledtabular}
\begin{table}[htb]
\caption{The distances between each quark pair in the configuration
of $D^*\bar{D}^*$ in different color structure. $r_{ij}$ stands for
$\sqrt{<r_{ij}^2>}$ ($j>i=1,2,3$) in this form. (unit: fm)
}\label{radius}
\begin{tabular}{cc|cccccc}
$J^{PC}$&Color structure &$r_{12}$&$r_{34}$ &$r_{13}$&$r_{24}$&$r_{14}$&$r_{23}$ \\
\hline
 &$1 \otimes 1$ & 0.65& 0.65&6.6 &6.6&6.6&6.7 \\
$0^{++}$ &$8 \otimes 8$ &0.7 &0.7 &0.69 &0.69 &0.36 &0.91\\
 & Coupling &0.71 &0.71 &0.69 &0.69  &0.36 &0.91\\\hline
 &$1 \otimes 1$ & 0.65& 0.65&6.63 &6.63&6.6&6.6 \\
$2^{++}$ &$8 \otimes 8$ &0.8 &0.8 &0.78 &0.78 &0.39 &1.03\\
 & Coupling &0.65 &0.65 &6.62 &6.62  &6.59 &6.64\\
\end{tabular}
\end{table}
\end{ruledtabular}
\end{center}

\section{Summary \label{summary}}
In summary, we have studied the possible molecular states of
$X(3915)$ and similar state $X_B$ with $b$ quark replacing c quark by
taking into consideration of the coupling of different color structures
in BCN model. In the quark model, there is no net interaction between
two color singlet mesons. To form the molecular state, the hidden color
channel must be introduced. By considering the hidden color channel
and the coupling between color singlet channel and the hidden color channel,
bounded states $D^*\bar{D}^*$ and $B^*\bar{B}^*$ are obtained.
In our calculation, The $X(3915)$ can be
interpret as $D^*\bar{D}^*$ with $J^{PC}=0^{++}$, which is in
agreement with the discussion in Ref.
\cite{CZYuan0910,yang:X3915,Liuxiang:Y3940epjc,Liuxiang:Y3940,Branz:Y3940,JianRongZhang:Y3940}.

Since the experimental data about the $X$(3915) is preliminary, further
confirmation by other experimental collaborations is necessary. The
role of the hidden color can be tested in multi-quark system if the
$X$(3915) is identified as a molecular state $D^*\bar{D}^*$ with
$J^{PC}=0^{++}$. At present, we only have a limited understanding
for the puzzling state $X$(3915). A systematical study of the
role of hidden color in different four-quark configurations will help
us to understand the effects of hidden color configurations. 
The work is in progress. 

\acknowledgments{We would like to thank Prof. E. Hiyama
for helpful discussion on GEM. This work is supported partly by the
National Science Foundation of China under Contract No.10775072, the
Research Fund for the Doctoral Program of Higher Education of China
under Grant No. 20070319007, No. 1243211601028 and Science
Foundation of Guizhou Provincial Eduction Department under Grant No.
20090054.}

\end{document}